# Towards improving the e-learning experience for deaf students: e-LUX

Fabrizio Borgia[1,2], Claudia S. Bianchini[3], and Maria De Marsico[2]

[1] Université Toulouse III - Paul Sabatier, 18 route de Narbonne, 31062 Toulouse Cedex 9, France
[2] Sapienza Università di Roma, Dip. Informatica, Via Salaria 113, 00198 Rome, Italy
fborgia, demarsicog@di.uniroma1.it
[3] Université de Poitiers, 1 rue Raymond Cantel (Bat A3 - UFR L&L), 86073 Poitiers Cedex 9, France
claudia.bianchini@univ-poitiers.fr

**Abstract**. Deaf people are more heavily affected by the digital divide than many would expect. Moreover, most accessibility guidelines addressing their needs just deal with captioning and audio-content transcription. However, this approach to the problem does not consider that deaf people have big troubles with vocal languages, even in their written form. At present, only a few organizations, like W3C, produced guidelines dealing with one of their most distinctive expressions: Sign Language (SL). SL is, in fact, the visual-gestural language used by many deaf people to communicate with each other. The present work aims at supporting e-learning user experience (e-LUX) for these specific users by enhancing the accessibility of content and container services. In particular, we propose preliminary solutions to tailor activities which can be more fruitful when performed in one's own "native" language, which for most deaf people, especially younger ones, is represented by national SL.

## 1 Introduction

Just like any other software, e-learning applications need to be understood by their users in order to be effectively exploited. In particular, they require an even more careful design, during which many issues must be addressed. Most of all, they must be so easy to use to become "transparent" and let the user focus on the true goal of their use, i.e. learning. A first point to take in consideration is the difference between the container, i.e. the software platform or framework to deploy the e-learning content, and the content itself, i.e. the learning material [1]. The former might be designed once and for all, while the latter undergoes continuous development and update. It is often the case that attention merely focuses on features available on the container, and this is also true when addressing accessibility issues. However, content, and particularly the way it is conceived and structured, is even more important when addressing users with special needs. This calls for devising ways to transmit the information through the best suited (sensory) channels for each category of users. Moreover, in many cases not only the transmission medium, but also the structure of the content must be adapted and deeply revised [2] [3].

Deaf people are among those groups of individuals with special needs who are most heavily affected by the digital divide. Despite the increasing attention towards accessibility issues, inclusion design for deaf people is often carried out using solutions on the edge of the workaround. In fact, most accessibility guidelines that should address their needs just deal with textual captioning and audio-content transcription [4]. However, this way of approaching the problem reveals an old misunderstanding: the cognitive structures underlying the language processing of deaf people are deeply different from those exploited by people experiencing Vocal Languages (VL) since infancy.

As a consequence, most deaf people have big troubles with VLs, even in their written form, so that the solution is many times not better than the problem.

While captioning-based accessibility design may support the needs of people who become deaf after the acquisition of speech and language (post-lingual deafness), issues related to pre-lingual deafness are seldom and poorly addressed. Moreover, all deaf people generally experience major difficulties with the written form of VLs, making their level of proficiency with it comparable to that of foreign students [5]. Among the possible reasons for this peculiarity, deaf people base their perception and cognitive structuring of information from exterior world mostly on the sense of sight. Therefore, they tend to reflect their visual organization of the world over the organization of language. For this reason, despite some deaf people can use VL, dealing with the issues of deaf-oriented accessibility using written VL is quite unrealistic, especially within the e-learning context.

Only a few organizations produced guidelines dealing with one of the most distinctive expressions of deaf people: Sign Languages (SL), i.e. the languages used by most deaf people in the world to communicate inside their community. Such languages are natural historical languages, based on the visual-gestural channel. The whole body (non just the hands, but even the arms, shoulders, torso, facial expression, and especially the look) is involved simultaneously and multimodally in the expression of the meaning (see the work of Pizzuto and Cuxac, such as [6] and [7]). The World Wide Web Consortium (W3C) has been one of the first organizations to require SL support in deaf-oriented accessibility design, through the Web Content Accessibility Guidelines (WCAG) 2.0 [8]. The WCAG 2.0 covers a wide range of recommendations for making Web content more accessible to a wide range of people with disabilities. Criterion 1.2.6, for instance, which is necessary to achieve the highest level of compliance (Level AAA), requires SL interpretation to be provided for all prerecorded audio content in synchronized media [8].

It is of paramount importance, for every group of individuals, to benefit from content written in one's own native language. Along this line, this work aims towards promoting e-learning accessibility for deaf people. In particular, we propose blueprints and preliminary solutions to support e-learning-tailored activities for deaf people, exploiting written SL.

## 2 Improving e-learning experience for deaf people

It is worth starting from the deep meaning that must be associated to the term "User Experience Design". It is quite easy to associate it to User Interaction Design, Information Architecture, Human-Computer Interaction, Human Factors Engineering, Usability and User Interface Design. As a matter of fact, according to [9], elements from each of these fields contribute to devising and implementing a positive user experience, and these fields overlap themselves. It is important to underline that accessibility has always been considered as one of the constituent facets of user experience [10]. It is not just an additional feature, it is a core component that makes modern interfaces complete. If designers fail to pay attention to the design needs for a small percentage of the population, they ultimately fail on a global scale" [11]. In this context, issues related to e-learning are especially ticklish. Given the problems often faced by people with special needs when interacting with ICT, the risk is that the kind of divide experienced in everyday life can also reect on a kind of even more critical digital divide. As a matter of fact, unique occasions for accessing information and services otherwise hindered may be definitively lost. And if it is accepted that usability of e-learning applications is especially important [1], it is crucial to recognize that designing effective e-learning containers and contents for users with special needs requires a wider group of competences [2]. In this scenario, experts and subjects possibly belonging to the target categories of users play a special role. This is long put in practice for some disabilities, especially blindness. Internet offers plenty of research, tools and services supporting blind or partially-sighted students, which, though not always effective, reveal a good understanding of their problems at least. The situation is dramatically different for deaf users, and more critically for their access to e-learning applications. Textual captioning and subtitling of videos, as well as audio-content transcription have been long considered as viable strategies to support these users [4]. However, this approach is grounded on a serious misunderstanding, based on the assumption that sight was able to play for deaf a symmetrical role to hearing for blind: "writing" an audio content to the former should have been as much effective as "speaking" a written text to the latter. While it is often the case even for blindness that changing the channel might not be sufficient in itself [2], it is further to consider that cognitive structures underlying the language processing of born deaf people are deeply different from those exploited by people experiencing VLs since infancy. As a consequence, most deaf people hardly handle VLs, even in their written form. Since the solution is many times not better than the problem, the quality of e-Learning User Experience (e-LUX) dramatically degrades. Both containers and contents must be carefully (re)designed.

## 3 Written SL and the digital world

A necessary condition to achieve the goal of full access for deaf learners to the digital world is integrating SL resources and tools within e-learning applications, since the benefits of this methodology cannot be matched by any other accessibility solution [4]. As a matter of fact, SL can be considered as a native language for a wider and wider part of the deaf community, especially younger ones, who did not experience the ostracism exercised by oralism supporters towards SLs. Oralism is based on the belief that full integration of deaf students requires their education to oral languages using among others tricks like lip reading, or mimicking mouth shape and breathing patterns of spoken speech. The decline of oralism is marked by the work by Stokoe [12], but not before the 1970s.

In deaf-oriented accessibility design, SL is primarily supported by including videos within digital artifacts. SL videos are usually employed to make a digital resource available to deaf people, but, in most cases, they are included into VL-featuring artifacts [4].

A few technologies have been developed to produce artifacts exclusively oriented to deaf-people, in order to provide both SL-based navigation and content fruition. Sign Language Scent [4] and SignLinking [13] are among those technologies. They all implement the Hypervideo pattern, which provides navigation and content fruition solutions by embedding one or more hyperlinks in SL videos, thus allowing the user to navigate and retrieve further information if interested in the concept conveyed by a particular sequence (i.e. at a particular time interval) in a SL video.

Though effective when exploitable, videos cannot completely substitute written text, most of all in tasks requiring continuous interaction with the system. While textual chat can be nowadays easily substituted by a video one, and annotation and tagging can be achieved by attaching short clips to the relevant contents, search for relevant information using SL is much harder to design and implement. This is a crucial task. Image processing and pattern recognition techniques applied to videos are computationally (more) demanding and may lack the required accuracy. A viable solution seems to be the adoption of a writing system able to transcribe SLs in an intuitive way.

Like most languages in the world, SLs have not developed, during the course of their long history, a writing system achieving a wide recognition within its community. However, unlike VLs without a written form, SLs cannot be represented through the adaptation of a pre-existing notation (e.g., the International Phonetic Alphabet) because of their nature and their visual-gestural multilinearity and simultaneity [14] [15]. Designing a writing system from scratch has therefore proven necessary to solve the problem of the graphical representation of SLs.

Different writing systems have been devised for SL over time: Stokoe's notation [12], in 1960, was the first one to achieve an international dissemination, followed, in the late 1970s, by SignWriting (SW) [16] and by Hamburg Sign Language Notation System (HamNoSys) [17], during the 1980s, see Fig. 1 for a visual comparison between the three systems[1].

---

[1] For a comparison between the three writing system, which is beyond the scope of this paper, please refer to [18]

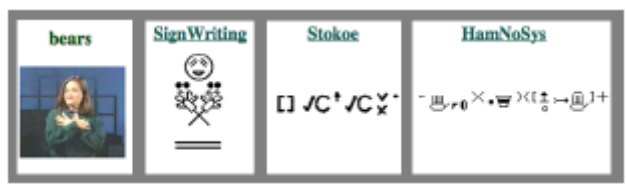

Fig. 1. ASL sign for "Bear", comparison between three different writing systems, extracted from [19]; from left to right: SW, Stokoe, HamNoSys.

Each of the aforementioned writing systems has its strengths and weaknesses, among them, SW proved more compatible with our goals, for reasons that are explained later in this section. According to Sutton [20], SignWriting is a writing system which uses visual symbols to represent the hand shapes, movements, and facial expressions of SLs. It is an "alphabet", a list of symbols used to write any SL in the world. Fig. 2 represents the Italian Sign Language (LIS) sign for "Fun", written in SW.

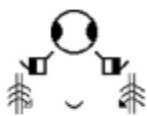

Fig. 2. LIS sign for "Fun", written in SW (extracted from [21])

Our reasons to adopt SW as the preferred SL writing system for our work are basically two: SW is a very iconic system, and it actually represents the physical production of the signs. The high iconicity of the system is due to the shapes of the symbols themselves, which are 2-dimensional abstract images depicting positions or movements of hands, face, and body. The spatial arrangement of the symbols on the page also plays a core role since it does not follow a sequential order (like the letters that make up written English words) but it follows the natural arrangement suggested by the human body.

As mentioned before, SW preserves the actual physical formation of the signs, using an analogy-based representation, but no information is conveyed about their meaning; no phonemic or semantic analysis of a language is required to write it. A person who has learned the system can "feel out" an unfamiliar sign in the same way an English speaking person can "sound out" an unfamiliar word written in the Latin alphabet, without even needing to know what the sign means" [19].

The set of movements and positions that a human body can produce from the waist up is huge. As a consequence, the set of symbols that SW provides to write down any sign is accordingly vast (approximately 38,000 units). The whole set of symbols is referred to as the International SignWriting Alphabet (ISWA). Within the ISWA, the symbols are organized in sets (categories) and subsets (groups), according to anatomic and production rules (see [22] for more details). For instance, ISWA Category 01 contains any symbol related to hand configurations. Category 01 includes different groups, one of them being ISWA Group 01, which contains any hand configuration featuring a single extended index finger. The organization introduced with the ISWA also allows to identify a unique 13-digit code for each symbol, each code provides information about which category and group its associated symbol belongs to.

The 2010 version of the ISWA (ISWA 2010) [22] is the most recent official one, and it has been adopted as the current standard by many research teams. In practice, we preferred to adopt the revised version of the ISWA by Bianchini [14].

Just like any other writing system, a SW text can be produced simply using pencil and paper. Despite this, SW has been oriented to digital production since its early years. The first digital editor for SW, named "SignWriter" was developed in 1986 by Richard Gleaves [23]. Since then, many applications have been produced by different teams, delivered in different ways, ranging from desktop to web applications. These applications can be considered as a class of software which basically provides the same functionalities, which are:
- pick or type a symbol in order to insert it on the sign composition area;
- manage the symbols on the sign composition area;
- save the sign in one or more formats.

In the last years, SW has also been exploited to narrow the effects of the digital divide on deaf people, in fact, a fair number of SW-based resources is emerging on the Internet, to provide SL deaf-accessible content. The produced artifacts include websites, blogs, online SW editors and mobile applications. A notable example is "The SignWriting Website" [19], which is the main SW research and dissemination portal providing content both in VL and SL (through video and SW). Another ambitious project is the ASL Wikipedia Project [24], whose goal is to provide an American Sign Language (ASL) Wikipedia written in SW. Finally, Adam Frost's "The Frost Village" [25] represents an example of bilingual blog, since it can be accessed both in VL (American) and SL (ASL). Witnessing the expressive capacity and the relative ease to learn of SW [14], our work focused on making it effectively exploitable as a communication mean, and as a suitable learning support for deaf people. To this aim, we designed, developed and tested a new SW digital editor, the SignWriting improved fast transcriber (SWift)[2] [3] [26]. SWift allows users to compose single signs or signed stories in a simple interface-assisted way, and save them in multiple formats. In order to produce an editor specifically tailored to the needs of its target users, we worked with the advice of experts and deaf researchers from the Institute of Cognitive Sciences and Technologies of the Italian National Research Council (ISTC-CNR), following the principles of contextual design [27]. The software has been developed as a self-containing web application, therefore it can be easily included within a Document Object Model (DOM) element, thus allowing SWift to be included within any web-based learning platform, in order to provide a prompt SL support for e-learning. Actually, the editor has already been successfully integrated within a deaf-centered e-learning environment (DELE) [28]. Once embedded within DELE, SWift has been employed to provide language support both to authoring and to communication tools for learners, such as chat, and forums.

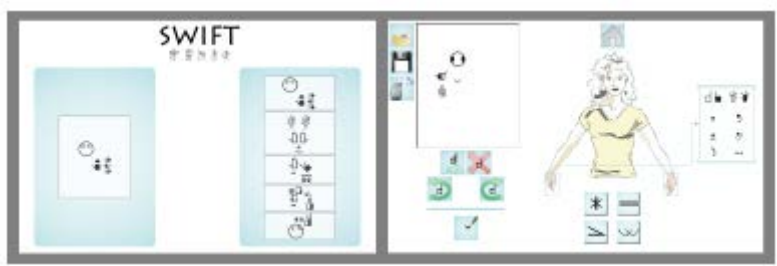

Fig. 3. Two screenshots taken from SWift. The home screen (left side) allows the user to choose between two modalities, depending on whether he wants to compose a single sign, or a whole signed story. The composition interface (right side) allows the user to actually enter a sign, in this case, the composition of the LIS sign for "Fun", shown in Fig. 2, is underway.

## 4 SignWriting Optical Glyph Recognition (SW-OGR)

Despite their increasing capabilities in terms of speed, usability, and reliability, SW digital editors are still far from granting the user a sign composition interface capable of challenging the simplicity of the non-digital handwritten approach. Actually, any software solution developed to support this need relies heavily on "Windows, Icons, Menus, Pointer" (WIMP) interfaces, both for accessing the application features and for the SW production process itself. The problem is all but a theoretical one, since in fact, during years of work along the ISTC-CNR team, which includes SW-proficient deaf researchers, we grew aware of a trend. We observed that SW users are far more accurate, fast and comfortable when using the traditional paper-pencil approach, than when dealing with the (more or less) complex interaction style of a digital editor.

Observing the intrinsic shortcomings of the present digital SW editors, we evaluated the possibility to design a new generation of SW editing applications, able to partially overcome the concept of the WIMP interface and to move along the line of the so called "natural interfaces" [29]. The new tools are intended to lift the user of any burden related to clicking, dragging, searching, browsing on the UI during the SW production process, and to provide him an interaction style which is as similar as possible the paper-pencil approach that humans normally use when writing or drawing. Of course, since WIMP is currently the easiest, most common interface style in the world, it cannot be totally left behind, because it is necessary to access the features of most applications. Nevertheless, our aim is to limit or dismiss the WIMP style during the SW production process, which is the core part of any SW editor.

---

[2] SWift is available at http://visel.di.uniroma1.it/SWift

To achieve this goal, we produced a SignWriting Optical Glyph Recognition (SW-OGR) engine, designed to operate the electronic conversion (recognition) of user-produced images containing handwritten (or printed) SignWriting symbols into machine-encoded (ISWA) SW text [30]. In particular. introducing a SW-OGR engine within an existing SW editor, such as SWift, will allow the user to handwrite symbols on the composition area, rather than searching them among thousands other symbols, and inserting them. The SW-OGR engine will operate the digital conversion of the handwritten symbols in real-time, or at the end of the composition process.

The main difference between OCR and OGR is in the nature of SW: it is composed by an extremely high number of elementary components (about 38,000 glyphs, compared with 26 letters of Latin alphabet) in a multilinear/two-dimensional arrangement, without rigid rules for such composition and, when handwritten, with a complex segmentation. OGR must be able to process such data, which is much more complex than alphabetic writing. Due to the huge amount of patterns to recognize, and consequently to the overwhelming training required, it is not feasible to exploit traditional pattern matching or machine learning approaches. We rather devised a software procedure driven by geometric as well as perceptual features of glyphs to recognize.

Our idea for the new generation of SW digital editors is illustrated by the diagram in Fig. 4. Our OGR-powered SW-editor is composed by the following modules:
- The Data Acquisition Module, included within the User Interface, whose purpose is to provide the user with a simple interaction style for SW composition, focusing on intuitiveness (or, better, invisibility) and accuracy. An ideal setting to provide a paper-pencil-like tangible interaction style requires the usage of an additional hardware component: the graphic tablet. However, a feasible alternative, though not always providing the sufficient stroke accuracy, is represented by touch screens. This module must also collect the data produced by the user (typically an image) and hand it to the SW-OGR module.
- The SW-OGR Engine, which globally handles and controls the recognition process. Two modules belong within the SW-OGR Engine:
  - the OGR Module, whose purpose is to provide a fast and as accurate as possible recognition of all (or most) symbols composed by the user;
  - the OGR Data Embedding Module, which is responsible for the creation of a result image embedded with the data produced by the OGR Module (it will most likely produce an image and an associated OGR data file, typically a SWML-encoded file).
- The data from the SW-OGR Engine are sent back to the User Interface, and are shown within the Review Module, which also allows the user to make corrections and/or add other data. The work done to produce SWift could prove very useful during the development of this module, since their functionalities (symbol search and editing) are very similar.
- he Data Finalization Module which receives the user-reviewed OGR data from the User Interface. The purpose of this module is to save in the proper form (file, DB, etc.) the data it receives.

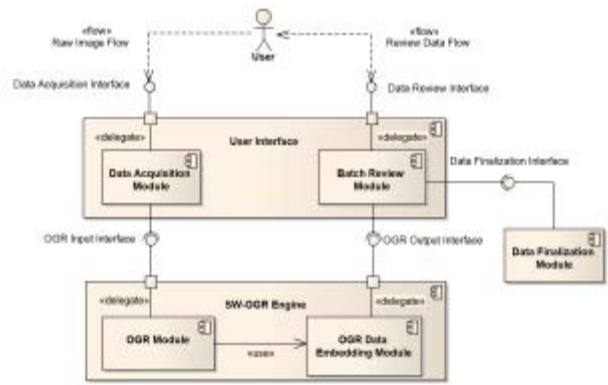

Fig. 4. Component diagram for a new generation of SW editors featuring a SW-OGR engine.

The achievement of such an editor is the main reason why we decided to work on a SW-OGR engine[3], since we think that it is bound to be the core component of a future SW digital editor (as illustrated in Fig. 4).

Another motivation fueled our efforts while working on SW-OGR. Since the beginning of our work, we were aware of the presence (and of the considerable size) of a number of handwritten SW corpora, produced from different communities around the world. In [14], for instance, Bianchini gathered every story written using SW by the deaf community of ISTC-CNR. Such corpora are an invaluable asset, and they could become even more useful if digitalized. Besides allowing a wider and faster diffusion of the information carried by the corpora themselves, we expect such resources to have several fields of application. They could, for instance, provide whole new SW datasets for the linguistic research community to perform any kind of analysis. Moreover, the produced digital SW documents could be employed together with 3D signing avatars. These kind of avatars are designed to convert entered (VL) text into SL [31]. Our opinion is that they could be adapted by implementing a SW input feed. Such methodology would improve the accuracy of the 3D avatars and make them more understandable by deaf users. In fact, the input feed would better resemble pure SL, while preconverting it in VL would impoverish its expressive richness. Furthermore, ISWA codes may provide finer production directives for the movements of the avatars.

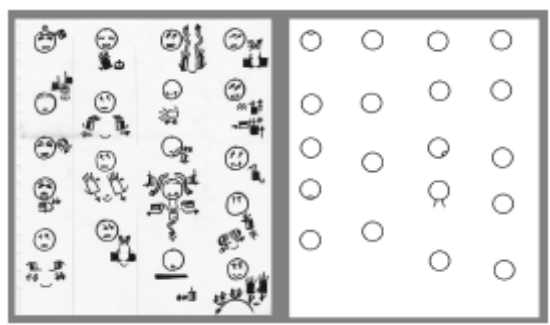

Fig. 5. Screenshot taken from SW-OGR during the development phase (ongoing).
The image shows a handwritten text (courtesy of ISTC-CNR) and
the result of the detection routine for the head symbols.

Since the SW-OGR provides very fast recognition routines for a large number of SW symbols (see Fig. 5), it can be employed both in real-time (e.g., SW editing) and batch (e.g., mass SW corpora digitalization) processing. In the second case, the diagram in Fig. 4 can be easily adapted to forbear the constant presence of a human actor (see [30]).

---

[3] For further information about SW-OGR, please refer to [30]

# 5 Conclusion and future

The inclusion of deaf people in the context of the best e-LUX achievements requires to carefully rethink the formulation and design of appropriate frameworks for ICT deployment. While SLs represent a widely spread form of expression for deaf people, their inclusion in electronic services and contents is still limited, when available, to videos. However, interactive tasks and search may take great advantage from exploiting a written system to translate SL expressions and contents in an intuitive way. This work is a brick in the creation of an overall framework targeting at supporting deaf people through their most natural form of expression. The final result, namely SWORD (SignWriting Oriented Resources for the Deaf) will represent a step towards full integration of deaf people in digital society.